\documentclass[twocolumn,amsmath,amssymb,superscriptaddress,fp]{jpsj3}
\usepackage{txfonts,amsmath,amssymb}
\usepackage{bm}
\usepackage{braket}
\usepackage{lipsum}
\usepackage{color}

\title{Nodal lines and boundary modes in topological Dirac semimetals with magnetism}
\author{Yasufumi Araki$^1$, Jin Watanabe$^2$, and Kentaro Nomura$^{2,3}$}

\inst{
    $^1$Advanced Science Research Center, Japan Atomic Energy Agency, Tokai 319-1195, Japan\\
    $^2$Institute for Materials Research, Tohoku University, Sendai 980-8577, Japan\\
    $^3$Center for Spintronics Research Network, Tohoku University, Sendai 980-8577, Japan
}

\abst{
    Nodal-line semimetals with magnetic orders have been theoretically predicted and experimentally observed in only few compounds.
    We theoretically explore the electronic structure in bulk and boundary of such a magnetic nodal-line state by introducing magnetism in topological Dirac semimetal (TDSM).
    TDSMs, such as $\mathrm{Cd_3 As_2}$ and $\mathrm{Na_3 Bi}$,
    are characterized by a pair of spin-degenerate Dirac points protected by rotational symmetries of crystals.
    By introducing local magnetic moments coupled to the electron spins in the lattice model of TDSM,
    we show that the TDSM can turn into either a Weyl semimetal or a nodal-line semimetal,
    which is determined by the orbital \modify{dependence in} the exchange coupling
    and the direction of magnetization formed by the magnetic moments.
    In this magnetic nodal-line semimetal state,
    we find zero modes with drumhead-like band structure at the boundary
    that are characterized by the topological number of $\mathbb{Z}$.
    \modify{Those zero modes are numerically demonstrated} by introducing magnetic domain walls in the lattice model.
}

\newcommand{\sech}{\mathrm{sech}\,}
\newcommand{\modify}[1]{\textcolor{black}{#1}}

\begin{document}
\maketitle

\section{Introduction}
\label{introduction}

Topological semimetals are characterized by their nodal band structures,
namely the touching of valence and conduction bands,
in three-dimensional (3D) momentum space \cite{Murakami_2007,Burkov_2011_2,Bansil_2016,Burkov_2016}.
While Dirac and Weyl semimetals are characterized by point nodes with linear energy-momentum dispersion around them \cite{Wan_2011,Burkov_2011,Young_2012,Steinberg_2014,Armitage_2018,Burkov_2018},
band touching along a 1D curve in momentum space is also possible in some crystals,
which are called nodal-line semimetals \cite{Kim_2015,Fang_2015,Chan_2016,Yamakage_2016,Fang_2016}.
Band inversion is essential in realizing those nodal structures,
which is usually introduced by crystalline structure or spin-orbit coupling.
Both nodal points and nodal lines are associated with topological invariants in the bulk,
which have correspondence to low-dimensional localized states at the boundaries of the system.
Dirac and Weyl semimetals exhibit quasi-1D Fermi arc states on the surface \cite{Wan_2011},
whereas nodal-line semimetals show drumhead surface states with nearly flat bands \cite{Chan_2016,Yamakage_2016}.

It was predicted at the early stage of theoretical studies
that those point-node and line-node structures are possible in both cases with and without time-reversal symmetry \cite{Burkov_2011_2}.
The first experimental realizations of Dirac \cite{Borisenko_2014,Liu_2014,Neupane_2014}, Weyl \cite{Lv_2015,Lv_2015_2,Huang_2015,Xu_2015,Xu_2015_2,Xu_2015_3,Xu_2015_4}, and nodal-line semimetals \cite{Hu_2016,Schoop_2016,Bian_2016,Bian_2016_2,Takane_2016} were in nonmagnetic compounds with time-reversal symmetry.
After those findings,
a number of theoretical and experimental attempts have been made
to realize topological nodal structures coexisting with magnetic orders.
Weyl semimetals, with both ferromagnetic \cite{Liu_2018,Xu_2018,Wang_2018,Morali_2019,Chang_2016,Ma_2019,Sakai_2018,Guin_2019,Li_2020,Kono_2020} and antiferromagnetic \cite{Nakatsuji_2015,Nayak_2016,Kuroda_2017,Higo_2018} orders, were experimentally synthesized with several faimilies of magnetic compounds,
which are now being intensely measured \cite{Manna_2018}.
On the other hand, the materials predicted as candidates for nodal-line semimetals with magnetism are still limited \cite{Jin_2017,Kim_2018,Nie_2019,Chen_2019,He_2021},
and clear experimental evidence of magnetic nodal lines is not yet established.
Moreover, it is still in question if such magnetic nodal lines in the bulk have any correspondence to boundary modes,
similarly to the drumhead surface states of nonmagnetic nodal-line semimetals.
We hence need to systematically understand the relations among magnetic orders,
nodal structure in the bulk, and the structure of boundary modes.

\modify{Regarding} the aforementioned background,
we here aim to understand the characteristics of such a magnetic nodal-line state
by introducing magnetization in nonmagnetic topological semimetals,
in particular the model of topological Dirac semimetal (TDSM) as the starting point.
TDSM is characterized by a pair of spin-degenerate Dirac points protected by rotational symmetries of crystal \cite{Yang_2014,Burkov_2016_2,Taguchi_2020},
and is realized in $\mathrm{Na_3 Bi}$ \cite{Wang_2012,Liu_2014}, $\mathrm{Cd_3 As_2}$ \cite{Wang_2013,Neupane_2014,Uchida_2017,Uchida_2019,Crassee_2018}, etc.
The Dirac points in TDSM are realized by the band inversion from spin-orbit coupling,
and are protected by crystalline symmetries.
Since the system contains multiple atomic orbitals characterized with different orbital and spin angular momenta,
one can expect that the interplay of spin-orbit coupling with magnetism may turn the spin-degenerate Dirac points into more complex nodal structures.

\modify{
    There are several ways to introduce magnetism in TDSM.
    One way is to dope magnetic elements in the material,
    which may possibly generate ferromagnetic orders
    in a manner similar to dilute magnetic semiconductors \cite{Ohno_1998,Dietl_2010}.
    Magnetic heterostructures are also of interest,
    e.g., an transport measurement was performed with the thin-film heterostructure of $\mathrm{Cd_3 As_2}$ and ferromagnetic insulator \cite{Uchida_2019},
}
and there are also several theoretical proposals focusing on the spin-helical surface states \cite{Misawa_2019,Kobayashi_2020,Araki_2021}.
Moreover, some theoretical calculations and transport measurements suggest that
the ferromagnetic Weyl semimetal $\mathrm{Co_3 Sn_2 S_2}$ may \modify{also turn into a paramagnetic TDSM},
by changing the carrier density with atomic substitution \cite{Weihrich_2006,Kurodera_2006,Kassem_2015,Kassem_2016,Thakur_2020}.

\modify{
    To consider the effect of magnetism on the nodal structure in those systems of TDSM,
    we here employ
}
the lattice model of TDSM coupled with local magnetic moments
\modify{
    and calculate
}
the low-energy band structure under the magnetization.
\modify{
    As a result,
}
we find that both Weyl-point and nodal-line structures are possible \cite{Burkov_2011_2}.
\modify{
    The nodal structure is governed by both
    the magnetization direction and the orbital dependence in the exchange coupling,
    which we will summarize in the form of topological phase diagram.
}
Furthermore, we point out the emergence of boundary modes in the \modify{obtained} magnetic nodal-line phase.
From the lattice-model calculation with magnetic domain wall texture,
we find localized states at the domain wall with zero-energy flat bands
\modify{
    enclosed by the nodal rings,
    similarly to the drumhead states on the surfaces.
}
The localized states in the present magnetic nodal-line phase are characterized by the topological number of $\mathbb{Z}$,
unlike $\mathbb{Z}_2$ topological number in nodal-line semimetals protected by mirror symmetry \cite{Kim_2015,Fang_2015,Chan_2016,Yamakage_2016}.
Those localized states contribute to the electric charging of domain walls,
which we show by evaluating the charge distribution on the lattice model.

This article is organized as follows.
In Section \ref{sec:model}, we introduce the low-energy effective model of TDSM in continuum and on lattice,
and define the possible structure of the exchange coupling term.
In Section \ref{sec:nodal-structures},
we calculate the band structure under the uniform magnetization,
and classify the obtained nodal structures in the form of topological phase diagram.
In Section \ref{sec:domain-wall},
we focus on the magnetic nodal-line phase,
and discuss the eigenstate structure under magnetic domain walls.
Finally, in Section \ref{sec:conclusion},
we summarize our findings and conclude our discussions.
We use the natural unit $\hbar=1$ throughout this article.

\section{Model Hamiltonian of TDSM with exchange coupling}
\label{sec:model}

In this section,
we introduce a model Hamiltonian of a TDSM
and consider the possible structure of the exchange interaction between the electron spins and the local magnetic moments.
A minimal model of a TDSM with a pair of Dirac points at low energy is derived from the $\boldsymbol{k} \cdot \boldsymbol{p}$ Hamiltonian as \cite{Wang_2012,Wang_2013}
\begin{align}
    H_0(\boldsymbol{k}) &= v_{xy}(k_x \tau_x \sigma_z - k_y \tau_y) +m(\boldsymbol{k}) \tau_z, \label{eq:continuum}
\end{align}
with $m(\boldsymbol{k}) = -m_0 + m_1 |\boldsymbol{k}|^2$.
This model consists of twofold spin times twofold orbital degrees of freedom,
and the Pauli matrices $\boldsymbol{\sigma}$ and $\boldsymbol{\tau}$ act on these two kinds of degrees of freedom, respectively.
In $\mathrm{Cd_3 As_2}$, for example, the basis of this Hamiltonian is composed of
\begin{align}
    \bigl\{ & |S_{1/2},j_z = +\tfrac{1}{2}\rangle, \ |S_{1/2},j_z = -\tfrac{1}{2}\rangle, \label{eq:basis} \\
    & |P_{3/2},j_z = +\tfrac{3}{2}\rangle, \ |P_{3/2},j_z = -\tfrac{3}{2}\rangle \bigr\}, \nonumber
\end{align}
which are from the $5s$ orbitals on Cd and $4p$ orbitals on As,
with $j_z$ the total angular momentum along the quantization axis $(z)$.
This model shows a pair of Dirac points $\bm{k}_{\mathrm{D}}^\pm = (0,0,\pm k_{\mathrm{D}})$ on $k_z$-axis,
with $k_{\mathrm{D}} = \sqrt{m_0 / m_1}$,
which are protected by the rotational symmetry around $z$-axis.
Around these Dirac points,
the energy bands show Dirac cones with the Fermi velocity $v_{xy}$ in $x$- and $y$-directions and $v_z = 2m_1 k_{\mathrm{D}}$ in $z$-direction.
On a hypothetical cubic lattice with the lattice constant $a$,
the tight-binding Hamiltonian
\begin{align}
    H_0^{\mathrm{lat}}(\bm{k}) &= t\left( \tau_x \sigma_z\sin k_x a - \tau_y \sin k_y a \right) + M(\bm{k}) \tau_z \label{eq:lattice-model} \\
    M(\bm{k}) &= -M_0 + 2M_1\sum_{i=x,y,z} \left(1-\cos k_i a \right),
\end{align}
reproduces the low-energy effective model [Eq.~(\ref{eq:continuum})] around $\boldsymbol{k} =0$,
with the Dirac points at $k_{\mathrm{D}} = a^{-1} \mathrm{arccos}(1-M_0/2M_1)$.
We shall use this lattice model,
with the parameters $M_0=t$ and $M_1 = 0.4t$, for the numerical calculations below.

Once we introduce magnetism in this system,
the local magnetic moments responsible for magnetism couple with the spins of the Dirac electrons.
If the magnetic moments are uniformly pointing in the direction of the unit vector $\boldsymbol{n}$,
the exchange coupling can be \modify{generally} written as
\begin{align}
    H_{\mathrm{exc}}[\boldsymbol{n}] &= \modify{ \sum_{i=x,y,z} \left(J_i \sigma_i + J'_i \tau_z \sigma_i \right) n_i } , \label{eq:exchange}
\end{align}
\modify{with six parameters $J_{x,y,z}$ and $J'_{x,y,z}$ characterizing the coupling structure.}
Since there are two orbital degrees of freedom in the present model,
the coupling structure may \modify{also} depend on the orbitals,
\modify{which is implemented by introducing two families of coupling parameters $J_{x,y,z}$ and $J'_{x,y,z}$.}

The \modify{coupling structure} should be restricted by the rotational symmetry originally present in the TDSM.
In particular, on the cubic lattice defined above,
the coupling structure needs to satisfy $C_4$-rotational symmetry around $z$-axis.
For the basis of $\mathrm{Cd_3 As_2}$ given in Eq.~(\ref{eq:basis}),
the $C_4$-rotation acts as the matrix
\begin{align}
    C_4 = \mathrm{diag} \left\{ e^{-i(\pi/2)j_z} \right\} = 
    \mathrm{diag} \left\{ e^{-i\frac{\pi}{4}}, e^{i\frac{\pi}{4}}, e^{-i\frac{3\pi}{4}}, e^{i\frac{3\pi}{4}} \right\},
\end{align}
under which the lattice Hamiltonian in Eq.~(\ref{eq:lattice-model}) is symmetric, $C_4 H_0(k_x, k_y, k_z) C_4^{-1} = H_0(-k_y, k_x, k_z)$.
We require that the structure of the exchange coupling term [Eq.~(\ref{eq:exchange})] should also be invariant under $C_4$,
\begin{align}
    C_4 H_{\mathrm{exc}}[n_x, n_y, n_z] C_4^{-1} = H_{\mathrm{exc}}[-n_y, n_x, n_z].
\end{align}
\modify{
    Since $C_4$ transforms the in-plane spin operators as
    \begin{align}
        \sigma_x \rightarrow \tau_z \sigma_y, \quad
        \sigma_y \rightarrow -\tau_z \sigma_x, \quad
        \tau_z \sigma_x \rightarrow \sigma_y, \quad
        \tau_z \sigma_y \rightarrow -\sigma_x,
    \end{align}
    $C_4$ rotational symmetry imposes the relations
    \begin{align}
        J_x = J'_y,
        \quad J'_x = J_y,
    \end{align}
    on the coupling parameters.
}
From this discussion, we obtain the general form for the exchange coupling term as
\begin{align}
    H_{\mathrm{exc}}[\boldsymbol{n}] &= J_{xy}(n_x \sigma_x + n_y \tau_z \sigma_y) + J_z n_z \sigma_z \\
    & \quad + J'_{xy}(n_x \tau_z\sigma_x + n_y \sigma_y) + J'_z n_z \tau_z\sigma_z, \nonumber
\end{align}
which is governed by four parameters $J_{xy},J_z$ and $J'_{xy},J'_z$.
For simplicity of discussion, here we require the isotropic coupling by taking $J_{xy} = J_z \equiv J$ and $J'_{xy} = J'_z \equiv J'$.

\section{Magnetization and nodal structures} \label{sec:nodal-structures}

\begin{figure}[b]
    \centering
    \includegraphics[width=8.4cm]{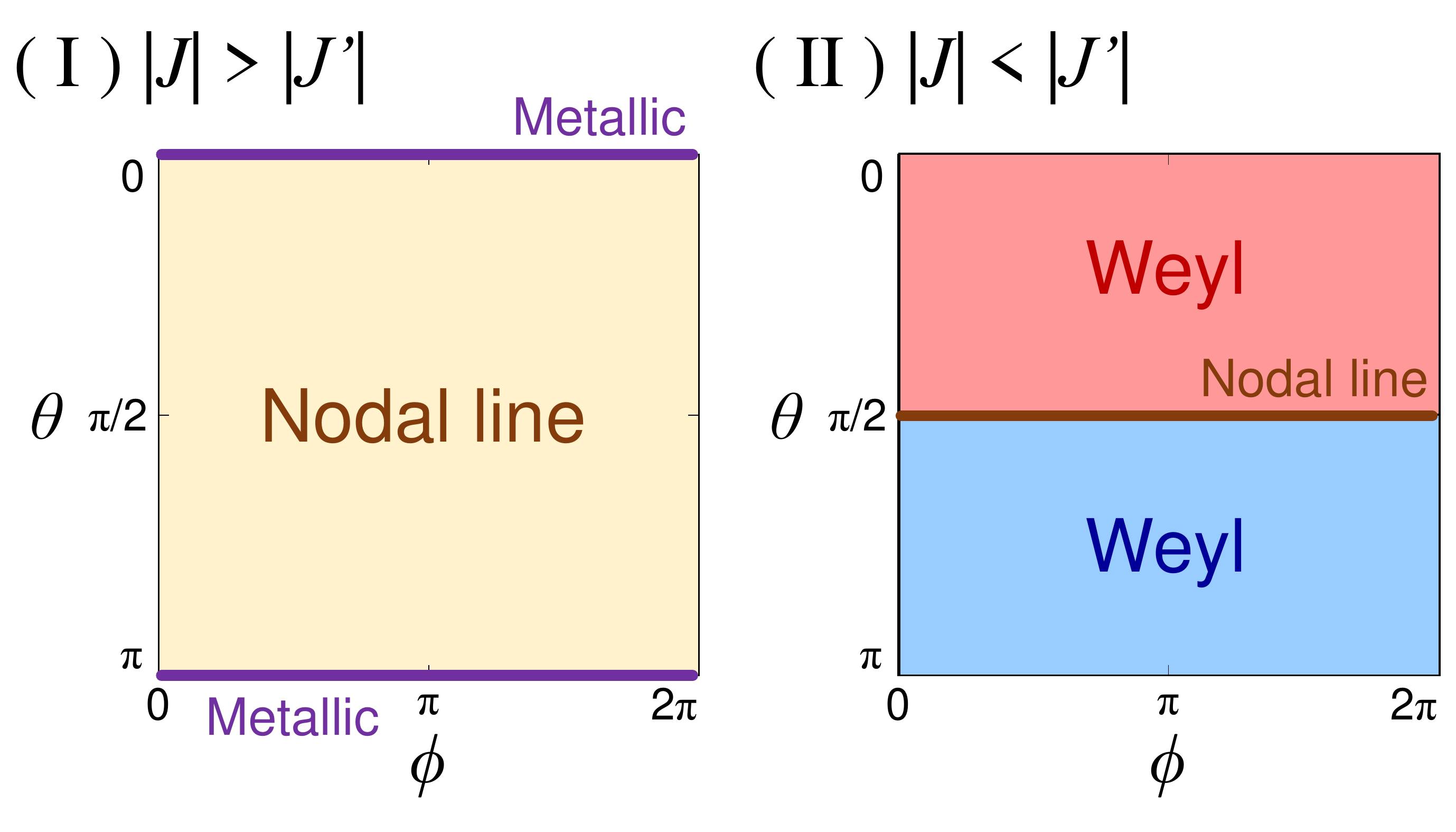}
    \caption{Schematic phase diagram of the topological nodal structure,
    mapped by the magnetization direction $\boldsymbol{n}=(\sin\theta \cos\phi, \sin\theta\sin\phi, \cos\theta)$.
    The possible phase structure is classified into the cases (I) and (II),
    which are determined by the sizes of the coupling parameters $J$ and $J'$.}
    \label{fig:phase-diagram}
\end{figure}

With the model Hamiltonian defined in the previous section,
we now consider how the nodal structure is influenced by the \modify{structure of} exchange coupling \modify{and the magnetization direction}.
Here we require the magnetic moments to be uniformly distributed,
forming a magnetization in the direction
\begin{align}
    \boldsymbol{n} = (\sin\theta \cos\phi, \sin\theta\sin\phi, \cos\theta).
\end{align}
Since there are two possible types of exchange coupling terms characterized by $J$ and $J'$ in the present model,
the nodal structure is also classified by the ratio between $J$ and $J'$.
We first summarize our findings of the nodal structure
in the form of phase diagrams,
\modify{
    in a manner similar to the previous studies on magnetic topological insulators \cite{Ominato_2018,Ominato_2019}
}
(see Fig.~\ref{fig:phase-diagram}).
\begin{description}
    \item[(I)] For $|J|>|J'|$: If $\boldsymbol{n}$ is parallel to $z$-axis ($\theta =0$ or $\pi$),
    the Dirac points split to Weyl points residing at finite energies,
    and the system becomes metallic at zero energy, \modify{with two spin-degenerate Fermi surfaces}.
    Otherwise the pair of Dirac points turn into a pair of nodal rings at zero energy.
    \item[(II)] For $|J|<|J'|$: If $\boldsymbol{n}$ lies in $xy$-plane ($\theta = \pi/2$),
    the pair of Dirac points turn into a pair of nodal rings at zero energy.
    Otherwise they turn into four Weyl points residing at zero energy,
    and the system can be regarded as a Weyl semimetal.
\end{description}
Below we show the details of the nodal structure.

\subsection{Case (I): $|J|>|J'|$}

\begin{figure}[tb]
    \centering
    \includegraphics[width=8.4cm]{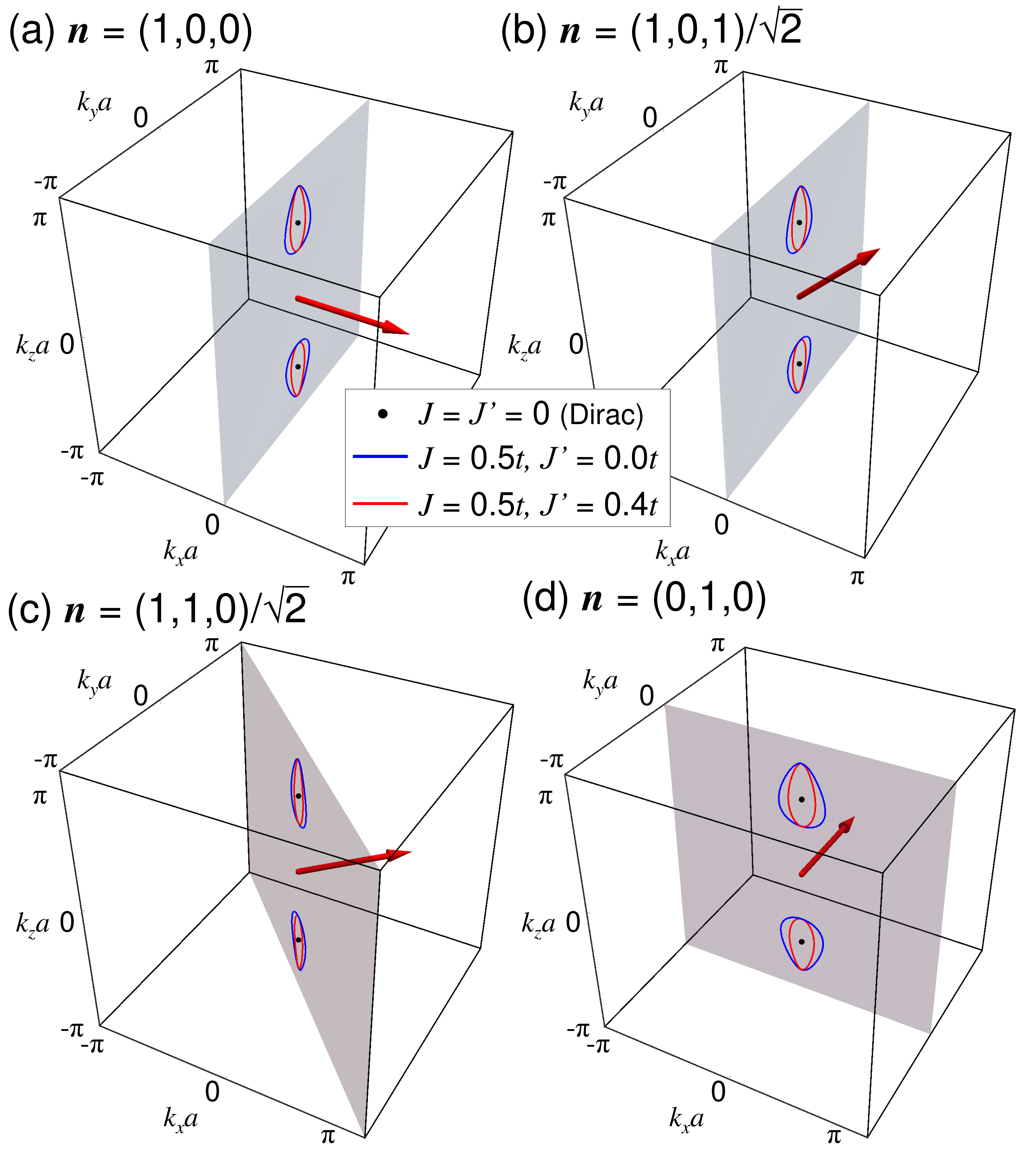}
    \caption{Structure of the nodal rings in momentum space for case (I) ($|J|>|J'|$),
    under the given direction of magnetization $\boldsymbol{n}$ and the exchange coupling parameters $(J,J')$.
    The red arrow in each panel denotes the direction of $\boldsymbol{n}$.
    The exchange coupling turns the Dirac points (black dots) into nodal rings (blue/red rings)
    for any direction of $\boldsymbol{n}$.
    The nodal rings reside on the gray plane on each panel.}
    \label{fig:nodal-rings}
\end{figure}

Let us first demonstrate the nodal structure with $J \neq 0$ and $J'=0$.
If the magnetization resides on $xz$-plane,
the energy eigenvalues of the net Hamiltonian $H(\boldsymbol{k}) = H_0(\boldsymbol{k}) + H_{\mathrm{exc}}$ can be analytically given as
\begin{align}
    & E(\boldsymbol{k}) = \label{eq:Ek-a} \\
    & \pm \left[ v_{xy}^2 k_x^2 n_x^2 +\left(\sqrt{v_{xy}^2(k_x^2 n_z^2 + k_y^2) + m^2(\boldsymbol{k})} \pm J\right)^2 \right]^{\tfrac{1}{2}}, \nonumber
\end{align}
which are doubly degenerate at $E=0$ for $\boldsymbol{k}$ satisfying
\begin{align}
    k_x n_x =0, \quad
    v_{xy}^2(k_x^2 n_z^2 + k_y^2) + m^2(\boldsymbol{k}) = J^2. \label{eq:nodal-ring-1}
\end{align}
If $\boldsymbol{n}$ has an in-plane component $(n_x \neq 0)$,
the double degeneracy occurs at the nodal rings satisfying $v_{xy}^2 k_y^2 + m^2(k_z^2) =J^2$ on the $k_x=0$ plane (Fig.~\ref{fig:nodal-rings}(a)(b)).
If $J$ is small enough,
we can linearize $m(\boldsymbol{k}) \approx \pm v_z (k_z \mp k_{\mathrm{D}})$ 
around the Dirac points $\boldsymbol{k}_{\mathrm{D}}^\pm$,
from which we obtain oval-shaped nodal rings
with the radii
\begin{align}
    R_{k_y} = J_0/v_{xy}, \ R_{k_z} = J_0/v_z \label{eq:radii}
\end{align}
along $k_y$- and $k_z$-axes, respectively.

Even if $\boldsymbol{n}$ is out of $xz$-plane,
the nodal rings can be seen
by rotating the band structure obtained above by the angle $\phi$ around $k_z$-axis (Fig.~\ref{fig:nodal-rings}(c)(d)).
On the other hand, if $\boldsymbol{n}$ is parallel to $z$-axis,
$\sigma_z$ serves as a good quantum number,
and hence the bands with $\sigma_z =+$ and $\sigma_z =-$ are just energetically split by $J$.
The Dirac point pair is split into two pairs of Weyl points,
with each pair residing at $E=\pm J$.
There arise spin-degenerate Fermi surfaces at $E=0$,
and the system becomes metallic.

The emergence of the nodal rings can be understood from the symmetry argument as follows.
As long as $J'=0$, the net Hamiltonian $H(\boldsymbol{k})$ possesses the chiral symmetry $\Gamma = \tau_x \sigma_y$,
which satisfies the anticommutation relation $\{ H(\boldsymbol{k}), \Gamma \} =0$.
The chiral symmetry requires this $4 \times 4$-Hamiltonian to have
two positive-energy eigenstates $|u^+_{1,2}(\boldsymbol{k})\rangle$
and two negative-energy eigenstates $|u^-_{1,2}(\boldsymbol{k})\rangle$,
which are related by $|u^-_{1,2}(\boldsymbol{k})\rangle = \Gamma |u^+_{1,2}(\boldsymbol{k})\rangle$.
Due to this relation, the Berry curvature for each band \cite{Thouless_1982,Kohmoto_1985,Haldane_1988},
which is defined by
\begin{align}
    \boldsymbol{\Omega}^\pm_n(\boldsymbol{k}) = i \langle \boldsymbol{\nabla}_k u^\pm_n(\boldsymbol{k}) | \times | \boldsymbol{\nabla}_k u^\pm_n(\boldsymbol{k}) \rangle, \ (n=1,2)
\end{align}
should satisfy $\boldsymbol{\Omega}^+_n(\boldsymbol{k}) = \boldsymbol{\Omega}^-_n(\boldsymbol{k})$ for any $\boldsymbol{k}$ in the Brillouin zone.
Since the Berry curvature summed over all the bands naturally vanishes,
i.e. $\sum_{n=1,2} [\boldsymbol{\Omega}^+_n(\boldsymbol{k}) + \boldsymbol{\Omega}^-_n(\boldsymbol{k})] =0$,
its sum over the negative-energy bands below $E=0$ also vanishes,
\begin{align}
    \sum_{n=1,2} \boldsymbol{\Omega}^-_n(\boldsymbol{k}) =0,
\end{align}
expect for degeneracy points.
Therefore, even if there arises a band \modify{touching} between the positive- and negative-energy bands at a certain $\boldsymbol{k}$,
it may not serve as a source or sink of the Berry curvature, namely a Weyl point,
and hence should form a nodal line or a degenerate Fermi surface residing at $E=0$ \cite{Burkov_2011_2}.

Even if the coupling parameter $J'$ satisfying $|J'| < |J|$ is also present,
we see the presence of the nodal rings by numerically diagonalizing the Hamiltonian,
displayed as the red rings in Fig.~\ref{fig:nodal-rings}.
In the presence of the coupling term with $J'$,
the nodal rings are generally lifted from $E=0$,
since this coupling term violates the chiral symmetry.
From the above calculations,
we have obtained the topological phase diagram shown in Fig.~\ref{fig:phase-diagram}(a),
with the metallic state for $\theta =0,\pi$
and the nodal-line state otherwise.

\subsection{Case (II): $|J|<|J'|$}

\begin{figure}[tb]
    \centering
    \includegraphics[width=8.4cm]{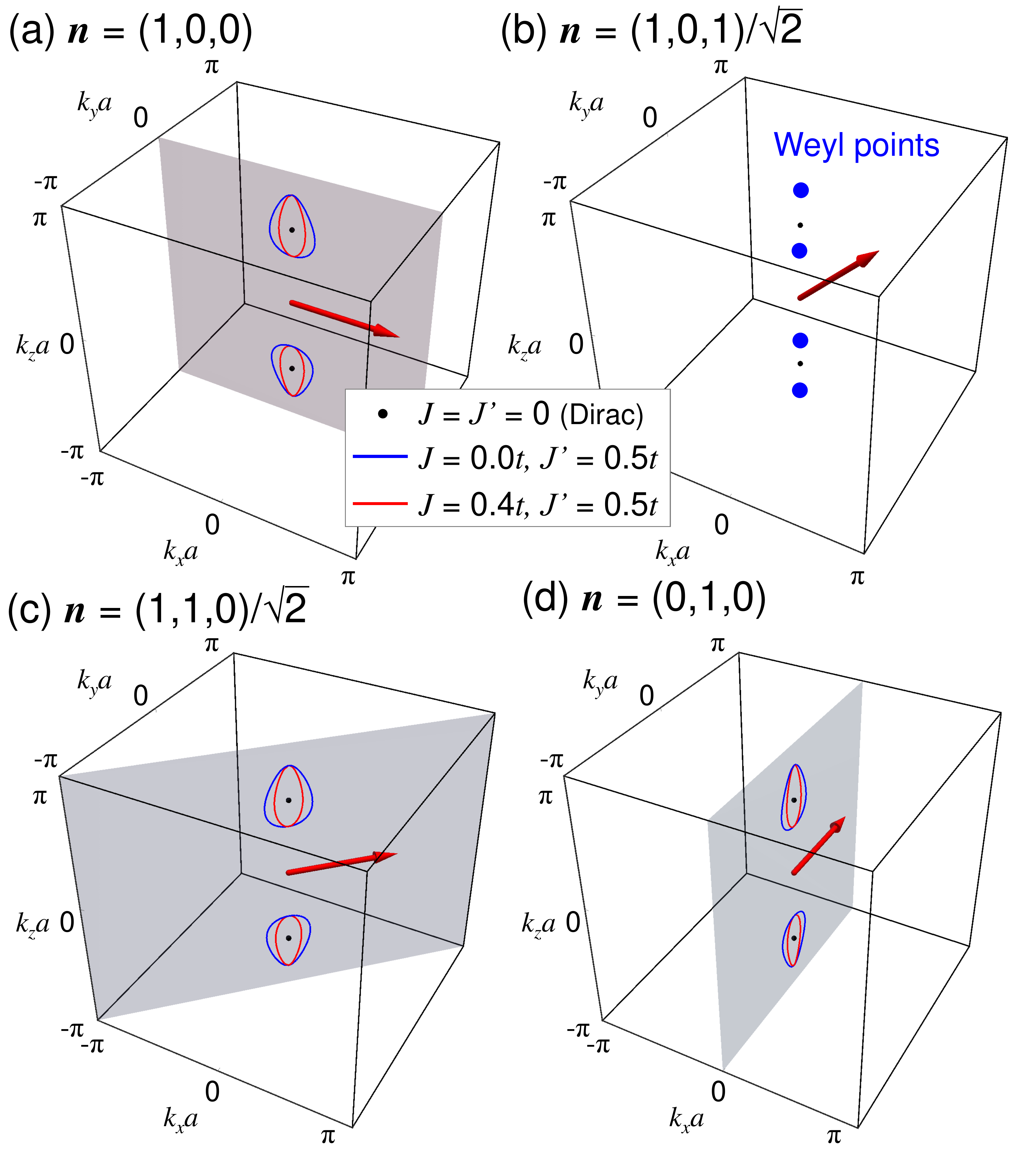}
    \caption{Structure of the nodal rings or the Weyl points in momentum space for case (II) ($|J|<|J'|$),
    under the given direction of magnetization $\boldsymbol{n}$ and the exchange coupling parameters $(J,J')$.
    The red arrow in each panel denotes the direction of $\boldsymbol{n}$.
    If $\boldsymbol{n}$ lies in $xy$-plane,
    the exchange coupling turns the Dirac points (black dots) into nodal rings (blue/red rings),
    which reside on the gray plane on each panel [(a)(c)(d)].
    In the presence of $n_z$-component,
    the Dirac points split into Weyl points (blue dots) residing on $k_z$-axis [panel (b)].
    }
    \label{fig:nodal-rings2}
\end{figure}

We again demonstrate the nodal structure first with $J' \neq 0$ and $J=0$.
If the magnetization $\boldsymbol{n}$ lies on $xz$-plane,
the energy eigenvalues of the net Hamiltonian $H(\boldsymbol{k}) = H_0(\boldsymbol{k}) + H_{\mathrm{exc}}$ can be analytically given as
\begin{align}
    & E(\boldsymbol{k}) = \label{eq:Ek-b} \\
    & \pm \left[ v_{xy}^2 (k_x^2 n_z^2 + k_y^2) +\left(\sqrt{v_{xy}^2 k_x^2 n_x^2 + m^2(\boldsymbol{k})} \pm J'\right)^2 \right]^{\tfrac{1}{2}}, \nonumber
\end{align}
which are doubly degenerate at $E=0$ for $\boldsymbol{k}$ satisfying
\begin{align}
    k_x n_z = k_y =0, \quad
    v_{xy}^2 k_x^2 n_x^2 + m^2(\boldsymbol{k}) = {J'}^2.
\end{align}
If $\boldsymbol{n}$ lies in $xy$-plane $(n_z =0)$,
the double degeneracy occurs at the nodal rings satisfying $v_{xy}^2 k_x^2 + m^2(\boldsymbol{k}) = {J'}^2$ on the $k_y=0$ plane.
The nodal rings are present for arbitrary direction of  $\boldsymbol{n}$ in $xy$-plane,
which can be seen by rotating the band structure obtained above by the angle $\phi$ around $k_z$-axis (see Fig.~\ref{fig:nodal-rings2}(a)(c)(d)).
On the other hand, if $\boldsymbol{n}$ has an out-of-plane component $(n_z \neq 0)$,
the double degeneracy occurs at four Weyl points satisfying $m(k_z^2) =\pm J'$ on $k_z$-axis, as shown in Fig.~\ref{fig:nodal-rings2}(b).
Here the Weyl points reside at $E=0$,
and hence the system can be regarded as a Weyl semimetal at $E=0$.
Even if the coupling parameter $J$ is present,
the nodal rings or the Weyl points are present as shown in Fig.~\ref{fig:nodal-rings2},
which are obtained by numerically diagonalizing the Hamiltonian.

\begin{figure}[tb]
    \centering
    \includegraphics[width=8.4cm]{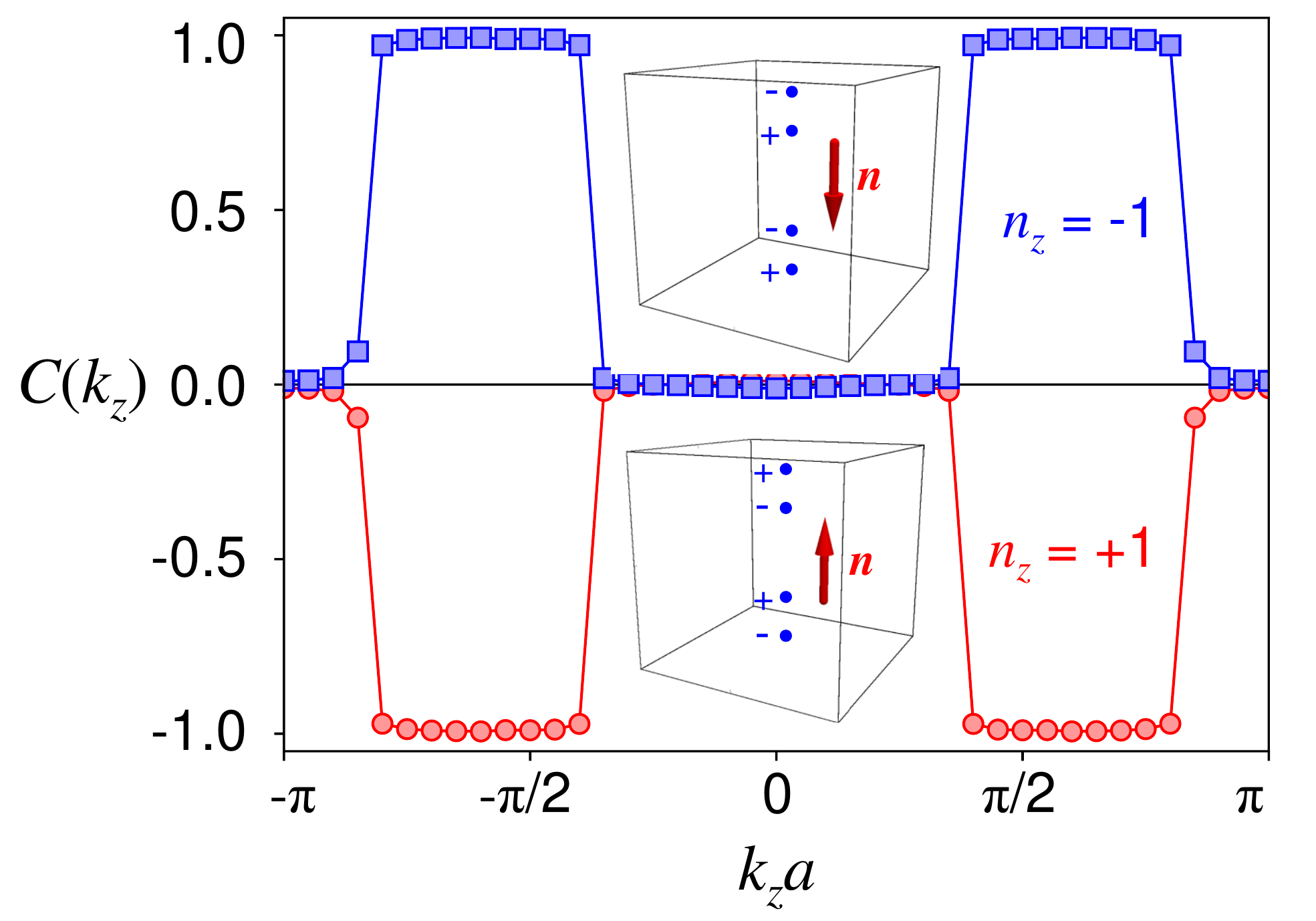}
    \caption{The Chern number $C(k_z)$ evaluated at fixed-$k_z$ planes in the lattice model,
    for the Weyl-semimetallic states with $J=0, J'=0.5t$, and $\boldsymbol{n}=(0,0,\pm 1)$.
    The inset shows the positions of the Weyl points and their topological charges for each case.}
    \label{fig:chern-number}
\end{figure}

In order to understand the characteristics of the Weyl-semimetallic state with $n_z \neq 0$,
we evaluate the topological charges of the Weyl points from the distribution of the Berry curvature.
From the Berry curvature $\boldsymbol{\Omega}_n(\boldsymbol{k})$ for each band $n$,
we here calculate the Chern number at fixed $k_z$,
\begin{align}
    C(k_z) &= \sum_{n \in \mathrm{occ}.} \int_{k_z} \frac{dk_x dk_y}{2\pi} \Omega_n^z(\boldsymbol{k}),
\end{align}
where the sum is taken over all the occupied bands below $E=0$.
$C(k_z)$ characterizes the Berry flux piercing through the 2D plane at this $k_z$,
and is related to the intrinsic anomalous Hall conductivity by $\sigma_{xy} = (e^2/4\pi^2) \int dk_z C(k_z)$ \cite{Thouless_1982,Kohmoto_1985,Haldane_1988}.
By using the lattice model defined in Eq.~(\ref{eq:lattice-model}) with $J=0$ and $J'=0.5t$,
we obtain the Chern number distribution $C(k_z)$ in the Weyl-semimetallic state
with the magnetization $\boldsymbol{n} \parallel \pm \boldsymbol{e}_z$,
as shown in Fig.~\ref{fig:chern-number}.
We can immediately see that the Chern number $C(k_z)$ takes a quantized value $\pm 1$ in some zones of $k_z$,
which correspond to the zones between the pair of Weyl points with opposite topological charges \cite{Burkov_2011,Burkov_2011_2,Armitage_2018}.
The sign of the Chern number, corresponding to the direction of the Berry flux,
is governed by the direction of the magnetization $\boldsymbol{n}$.
From this calculation result, we can identify the topological charges of the Weyl points as shown in the inset of Fig.~\ref{fig:chern-number},
whose signs are flipped depending on the sign of $n_z$.
These two states are topologically distinct;
if $\boldsymbol{n}$ is continuously varied from $+\boldsymbol{e}_z$ to $-\boldsymbol{e}_z$,
the gap between the Weyl points should close at a certain intermediate value of $\boldsymbol{n}$ to interchange the topological charges of  the Weyl points,
which corresponds to the nodal rings appearing at $n_z =0$.
As a result, we obtain the topological phase diagram as shown in Fig.~\ref{fig:phase-diagram}(b) mapped by the direction of $\boldsymbol{n}$,
which consists of the Weyl semimetallic phases distinguished by the topological charges of the Weyl points, and the nodal-line phase in between.

To summarize this section,
we have found that the magnetization coupled to the electrons in TDSM
can turn the Dirac points into \modify{either} Weyl points or nodal rings,
\modify{which is determined by} the direction of magnetization and the orbital \modify{dependence} in the exchange coupling.
We can thus expect TDSMs with magnetic dopants,
or TDSM thin films coupled with ferromagnets by the proximity effect,
as the good candidate for realizing the nodal-line semimetal state with broken time-reversal symmetry.

\section{Domain walls and localized states}
\label{sec:domain-wall}

So far we have seen the characteristics of the nodal structures in the bulk under a uniform magnetization.
In this section,
we consider how the nodal structure is related to the electronic properties at the boundaries,
particularly magnetic domain walls.
Based on the magnetic nodal-line state obtained in the previous section,
we numerically evaluate the eigenstate structure under a magnetic domain wall from the lattice model.
We demonstrate the emergence of drumhead in-gap states at the domain wall,
which appear quite similar to the well-known drumhead surface states.

\subsection{Lattice model}
We here construct a magnetic domain wall structure in TDSM.
Since we are interested in the behavior of the magnetic nodal-line state in this section,
we start with the case (I) obtained in the previous section,
by setting the parameters $J \neq 0$ and $J'=0$.
With those parameters,
the system shows the nodal lines unless the magnetization $\boldsymbol{n}$ is pointing exactly to $\pm z$-direction.
We here consider a head-to-head domain wall,
namely a boundary of two magnetic domains with the magnetizations $\boldsymbol{n} = \pm \boldsymbol{e}_x$ at the plain perpendicular to $x$-axis.
As seen in the previous section, each domain hosts a pair of nodal rings residing on $k_y k_z$-plane.
This head-to-head domain wall is likely to appear in a thin film geometry confined in $z$-direction,
since it minimizes the magnetostatic energy from the magnetic dipolar interaction.

\begin{figure}[tb]
    \begin{center}
    \includegraphics[width=8cm]{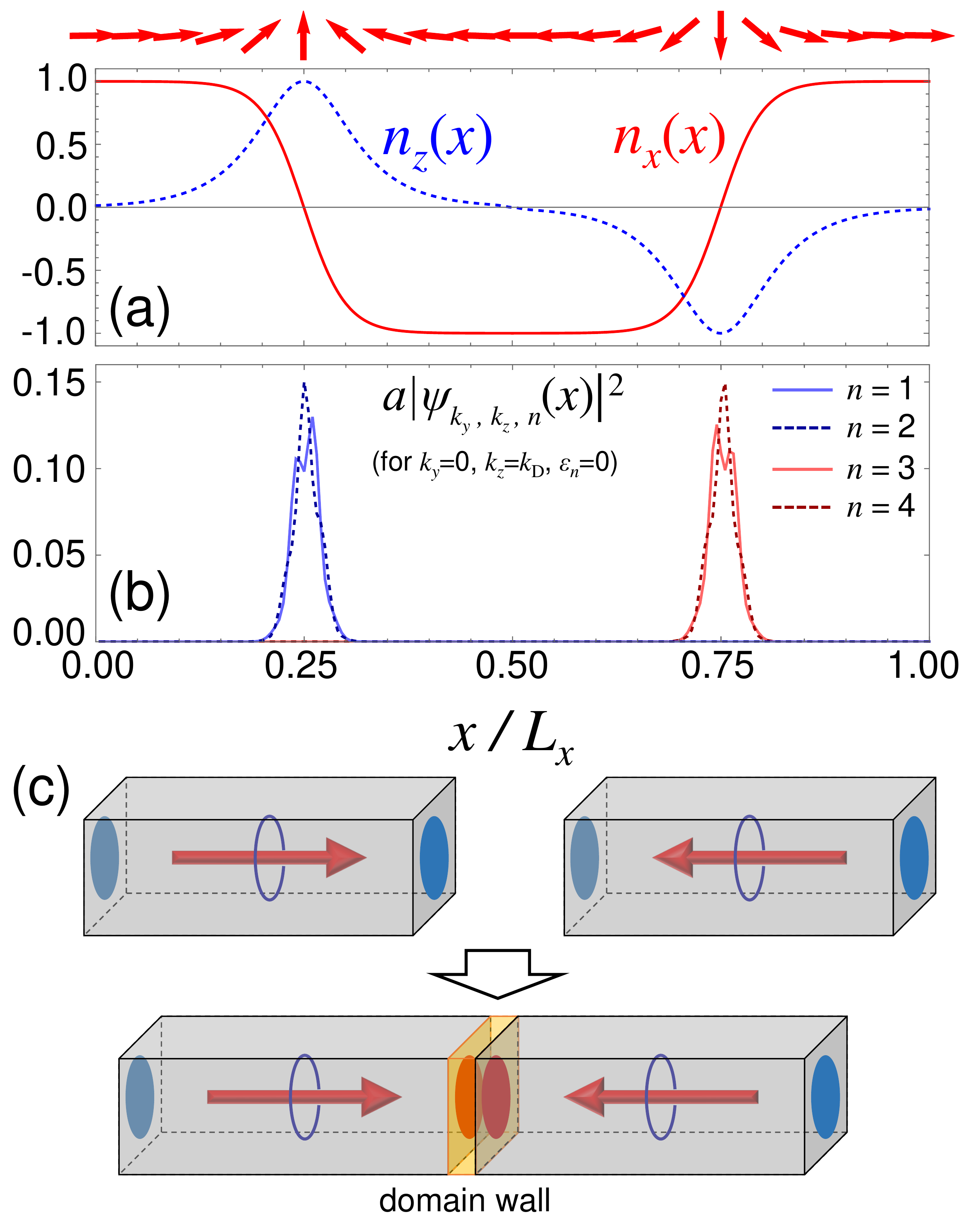}
    \caption{(a) The structure of the magnetic domain wall $\boldsymbol{n}(x)$ defined in Eq.~(\ref{eq:domain-wall}).
    We introduce head-to-head and tail-to-tail domain walls in the periodic boundary condition, as schematically displayed on the top.
    (b) The amplitude of the wave functions $|\psi_{k_y,k_z,n}(x)|^2$ under the domain wall texture,
    for the zero-energy modes (drumhead states) at $k_y=0, k_z = k_{\mathrm{D}}$.
    The index $n=1,\ldots 4$ is the label for the fourfold degenerate zero modes.
    The drumhead-state wave function is distributed mainly at the domain walls.
    (c) Schematic picture of the localized states obtained in the calculations.
    Drumhead states (blue discs) corresponding to the nodal rings emerge on the surfaces of each magnetic domain.
    Once two domains meet together by forming a head-to-head domain wall,
    twofold degenerate drumhead states (red discs) arise at the domain wall.}
    \label{fig:domainwall-wf}
    \end{center}
\end{figure}

For the numerical calculation, we need to implement this domain wall structure on the lattice model.
Based on the lattice model of TDSM defined by Eq.~(\ref{eq:lattice-model}),
here we treat its $x$-direction by the real-space formalism,
since the domain wall violates the lattice translational symmetry in this direction.
On the other hand, the transverse momentum components $k_y$ and $k_z$ are kept as good quantum numbers.
We impose periodic boundary conditions in all three dimensions,
with the lattice size $L_{x,y,z} ( = N_{x,y,z} a)$ for each direction.
Within this periodic boundary condition,
we need at least two domain walls in the system.
We thus formulate the domain wall structure as
\begin{align}
    \boldsymbol{n}(x) =
    \begin{cases}
        (-\tanh \xi_1(x), \ 0, \ \sech\xi_1(x)) & (0<x<L_x/2) \\
        (\tanh \xi_2(x), \ 0, \ -\sech\xi_2(x)), & (L_x/2 < x < L_x)
    \end{cases}
    \label{eq:domain-wall}
\end{align}
with
\begin{align}
    \xi_1(x) = \tfrac{1}{W}\left(x - \tfrac{L_x}{4}\right), \quad
    \xi_2(x) = \tfrac{1}{W}\left(x - \tfrac{3L_x}{4}\right)
\end{align}
for the domain walls of the size $W$ residing at $x=L_x/4$ and $x=3L_x/4$.
For the numerical calculations below,
we fix the system size $L_x = 200a, L_{y,z} = 100a$,
and the size of the domain walls $W = 10a$.
With those parameters, the magnetization $\boldsymbol{n}(x)$ takes the structure as shown in Fig.~\ref{fig:domainwall-wf}(a).

\subsection{Drumhead states at domain walls}

\begin{figure}[tb]
    \begin{center}
    \includegraphics[width=8.4cm]{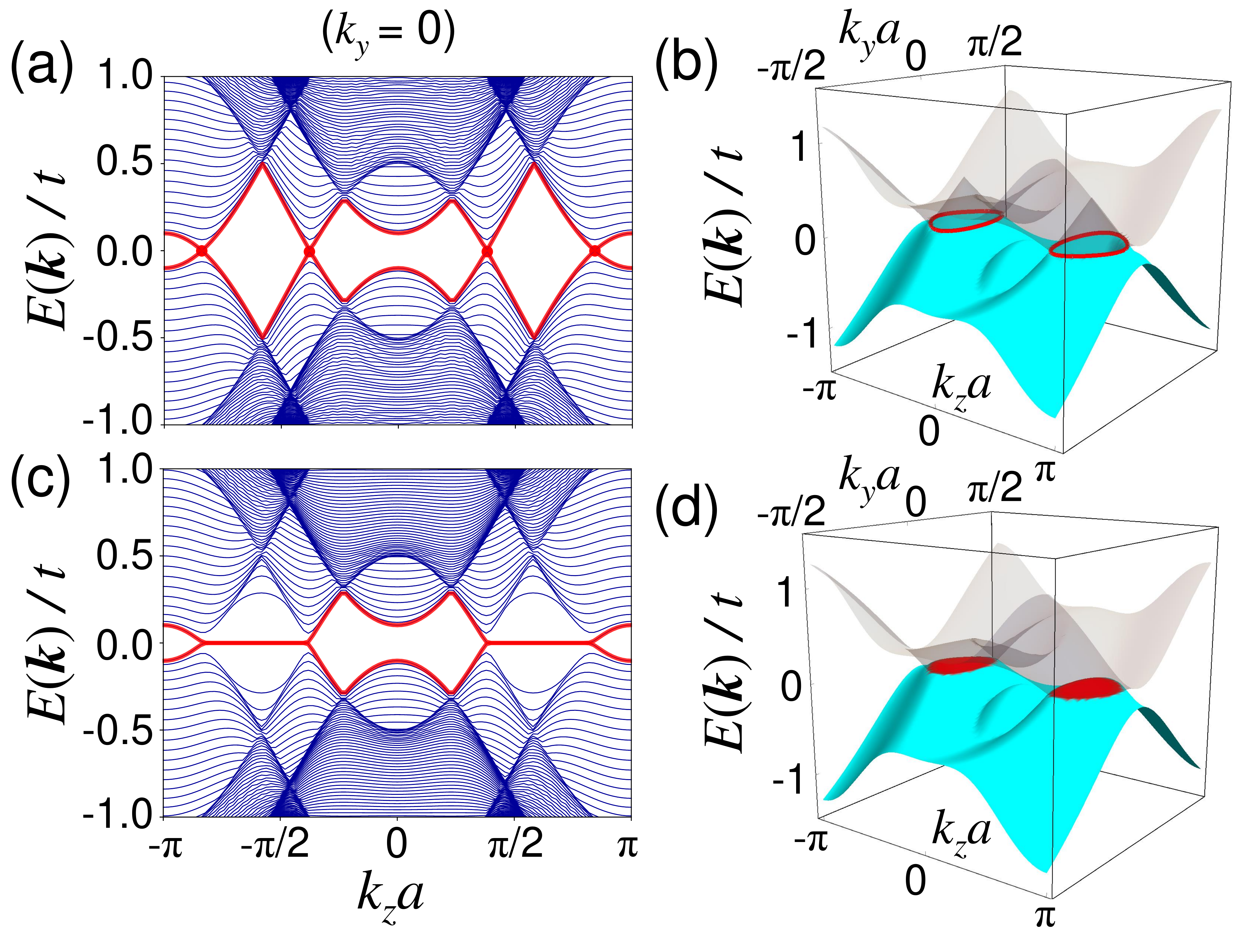}
    \caption{Band structure of the lattice model of TDSM 
    under uniform magnetization $\boldsymbol{n} = \boldsymbol{e}_x$ [(a)(b)],
    and that under the magnetic domain wall texture $\boldsymbol{n}(x)$ defined in Eq.~(\ref{eq:domain-wall}) [(c)(d)].
    The panels (a) and (c) show the slice of the bands at $k_y=0$.
    The panels (b) and (d) show two bands responsible for the band touching at zero energy,
    which are displayed as the red curves in (a) and (c).
    While the band touching occurs at the nodal rings under the uniform magnetization,
    it shows the flat drumhead structure in the presence of the domain wall.
    The parameters are taken as $J=0.5t$ and $J'=0$.}
    \label{fig:bands_domainwall}
    \end{center}
\end{figure}

By numerically diagonalizing the lattice Hamiltonian defined above,
we obtain the set of energy eigenvalues $\epsilon_{k_y,k_z,n}$
and eigenstate wave functions $\psi_{k_y,k_z,n}(x)$ for the given domain wall texture,
with $n(=1,\ldots N_x)$ the symbol for the band index.
While the band structure under the uniform magnetization $\boldsymbol{n}(x) = \boldsymbol{e}_x$ shows the nodal rings on $k_y k_z$-plane [see Fig.~\ref{fig:bands_domainwall}(a)(b)],
as discussed in the previous section,
the band structure under the domain wall defined in Eq.~(\ref{eq:domain-wall})
shows flat bands at zero energy,
which occur with fourfold degeneracy in the momentum region inside the nodal rings [see Fig.~\ref{fig:bands_domainwall}(c)(d)].
This flat-band structure
is similar to the drumhead states on the surface of nodal-line semimetals,
and hence we here call the obtained flat-band states as the drumhead states as well.

In order to understand the spatial structure of the drumhead states,
we plot the amplitude $|\psi_{k_y,k_z,n}(x)|^2$ of the obtained drumhead states at the momentum point $k_y=0, k_z = k_\mathrm{D}$,
which corresponds to the Dirac point in the nonmagnetic case,
in Fig.~\ref{fig:domainwall-wf}(b).
Among the fourfold degenerate drumhead states,
we can see that two states ($n=1,2$ in the figure) are localized around the domain wall at $x= L_x/4$,
and the other two states $(n=3,4)$ at $x = 3L_x/4$.
We may roughly understand the origin of such drumhead states at the domain walls
as the hybrid of drumhead surface states for the two topologically distinct magnetic domains,
as schematically shown in Fig.~\ref{fig:domainwall-wf}(c).
Such an idea on the drumhead states is similar to that on the ``Fermi arc'' states
found at magnetic domain walls in a Weyl semimetal \cite{Araki_2016,Araki_2018,Araki_2019}.
More precise argument on the drumhead states from the topological point of view is given later in this section.

\subsection{Localized charge at domain walls}
\begin{figure}[tb]
    \begin{center}
    \includegraphics[width=0.85\linewidth]{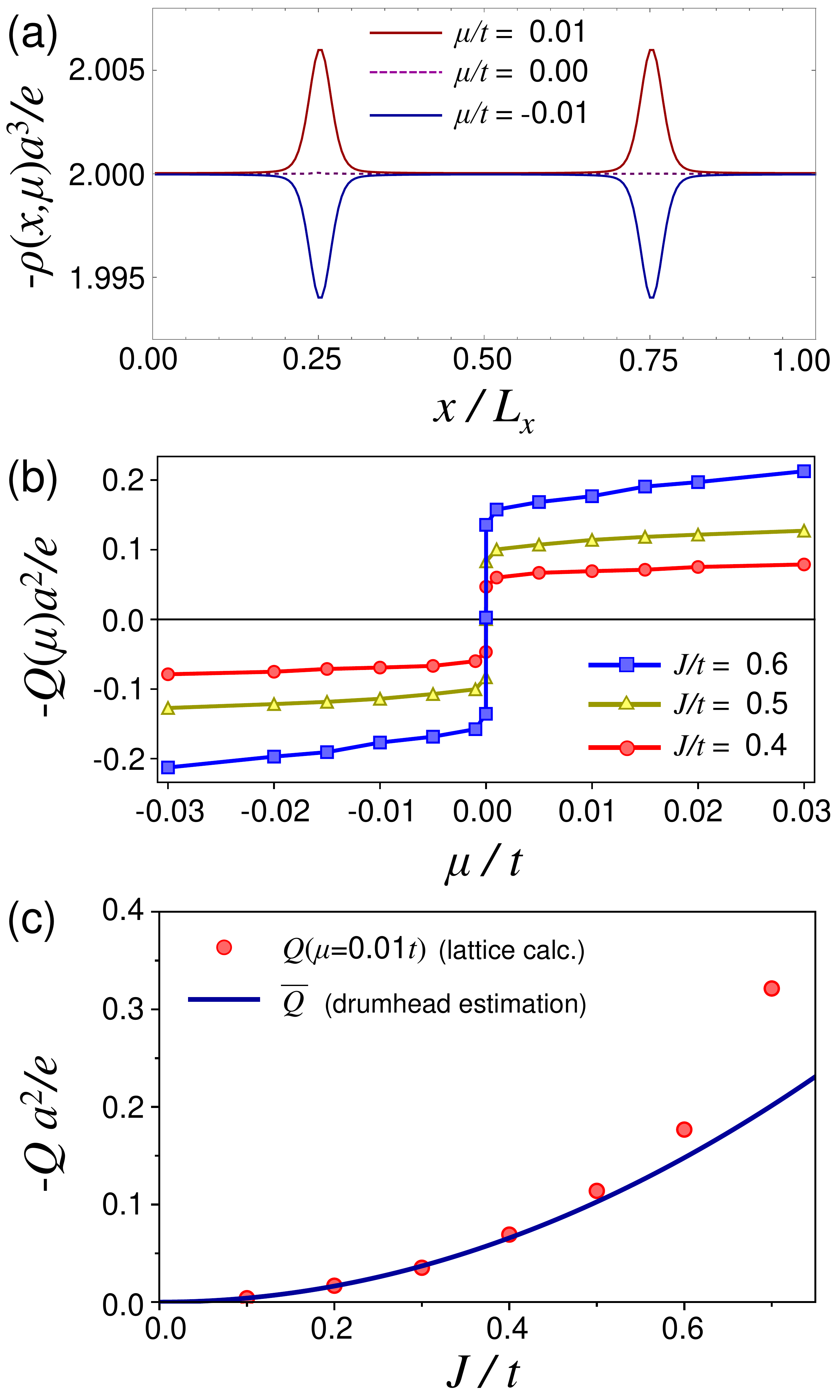}
    \caption{(a) The charge distribution $\rho(x,\mu)$ in the presence of the head-to-head domain wall
    calculated from Eq.~(\ref{eq:charge-distribution}).
    There arises localized charge around the domain walls at $x=L_x/4$ and $x= 3L_x/4$.
    (b) Behavior of the localized charge $Q(\mu)$ defined by Eq.~(\ref{eq:Q-mu}) as a function of the chemical potential $\mu$.
    The stepwise change at $\mu=0$ can be attributed to the zero-energy drumhead states arising from the domain wall texture.
    (c) Behavior of the localized charge $Q(\mu = 0.01 t)$ for several values of $J$.
    It agrees quite well with $\bar{Q}$ estimated from the area of the nodal rings [Eq.~(\ref{eq:approx-charge})] for small $J$.}
    \label{fig:domainwall-charge}
    \end{center}
\end{figure}

Since the drumhead states are localized at the domain walls and shows the flat band at zero energy,
they may give rise to nonuniform distribution of electric charge.
By fixing the chemical potential $\mu$,
the charge density distribution is given by the summation over all the occupied states below $\mu$ (at zero temperature),
\begin{align}
    \rho(x, \mu) = \frac{-e}{L_y L_z} \sum_{k_y,k_z,n} |\psi_{k_y,k_z,n}(x)|^2 \theta(\mu-\epsilon_{k_y,k_z,n}) , \label{eq:charge-distribution}
\end{align}
where each eigenstate wave function is normalized by $\int dx |\psi_{k_y,k_z,n}(x)|^2=1$.
The charge distribution around $\mu=0$ is shown in Fig.~\ref{fig:domainwall-charge}(a).
If $\mu$ is shifted slightly above zero energy,
the drumhead states are completely filled and hence the charge density shows peaks at the positions of domain walls,
whereas it shows dips for $\mu<0$ because the drumhead states are completely unoccupied.
If $\mu$ is set exactly at zero energy,
all the states in the system, including the drumhead states, are half-filled,
and hence the charge distribution becomes neutral and uniform.

From the charge distribution given above,
we can extract the net charge in the system measured in comparison with charge neutraity point,
\begin{equation}
    Q(\mu) = \int dx \left[\rho(x,\mu) - \rho(x, \mu = 0) \right], \label{eq:Q-mu}
\end{equation}
which is the 2D charge density per unit area in $yz$-plane.
The behavior of $Q(\mu)$ as a function of $\mu$ is shown in Fig.~\ref{fig:domainwall-charge}(b),
for several values of the exchange coupling $J$.
We can see a stepwise change in $Q(\mu)$ at $\mu=0$ (charge neutrality),
which comes from the large density of states
of the flat bands in the drumhead states localized at the domain walls.
This change becomes larger for large values of $J$,
since the size of the nodal rings is governed by $J$, as seen from Eq.~(\ref{eq:nodal-ring-1}).

In particular,
by using the approximate radii of nodal rings $(R_{k_y},R_{k_z})$ in Eq.~(\ref{eq:radii})
obtained from the linearized band structure,
we can expect that the localized charge at the domain walls is (almost) given from the area of nodal rings,
\begin{align}
    \bar{Q} = -4e \frac{\pi R_{k_y} R_{k_z}}{(2\pi)^2} = -e \frac{J^2}{\pi v_{xy} v_z}, \label{eq:approx-charge}
\end{align}
which is proportional to $J^2$.
Here the prefactor $4$ comes from the fourfold degeneracy of the drumhead states.
In order to check this $J$-dependence,
we plot in Fig.~\ref{fig:domainwall-charge}(c) the values of $Q(\mu)$ at $\mu = 0.01t$ calculated from the lattice model,
which we can mostly regard as the localized charge from the drumhead states,
and the approximate value $\bar{Q}$ given above.
We can see that $Q(\mu)$ coincides with $\bar{Q}$ quite well for $J \lesssim 0.5t$.
From these calculations,
we can understand that the drumhead states at the domain walls give
the dominant contribution to the sudden change in $Q(\mu)$ around charge neutrality,
which gives rise to the localized charge at the domain walls.

Several comments are in order
regarding experimental measurement of the localized charge in realistic materials.
First, we estimate how much charge may be localized at domain walls.
By using the Fermi velocities
$v_{xy} = 8.64 \mathrm{eV} \text{\AA}, \ v_z = 2.16 \mathrm{eV} \text{\AA}$ obtained from the measurements with angle-resolved photoemission microscopy (ARPES) in $\mathrm{Cd_3 As_2}$ \cite{Liu_2014},
the energy scale of the exchange coupling $J = 100 \mathrm{meV}$ obtained from first-principles theory calculations in magnetic topological insulators \cite{Yu_2010},
and the size of domain walls $W = 10 \mathrm{nm}$ as typical parameters,
we can estimate the charge density in the domain walls from Eq.~(\ref{eq:approx-charge}) as
\begin{align}
    \frac{\bar{Q}}{W} \approx -1.6 \times 10^{18} e \ \mathrm{cm^{-3}}.
\end{align}
This value is about ten times larger than that estimated in magnetic Weyl semimetals \cite{Araki_2018}.
Such a large difference in the localized charge comes from the dispersions of localized states at domain walls:
whereas the localized states in Weyl semimetals are the Fermi-arc states with quasi-1D dispersion,
those in the nodal-line semimetal obtained here are the drumhead states with the flat band structure,
which have a large density of states and contribute to such a large localized charge.

Another important effect to be considered is the electrostatic screening of this localized charge.
The electrostatic screening from the Coulomb interaction is inevitable
as long as the bulk density of states is finite,
which prevents us from detecting charged objects in metallic systems.
For nodal-line semimetals,
the drumhead states show the flat band structure at zero energy,
and the bulk density of states vanishes at the energy of the nodal rings,
if the bulk bands are particle-hole symmetric.
In this case, the localized charge arising from the zero modes is
almost free from the electrostatic screening by the bulk carriers,
and hence may be captured by the scanning tunnel microscopy (STM) in thin film geometry.
We should note that this is not the case in realistic $\mathrm{Cd_3 As_2}$,
which does not have particle-hole symmetry \cite{Wang_2013}.
In the absence of particle-hole symmetry, the bands of the drumhead states are not flat in general.
In order to make the drumhead states fully occupied, we should set the Fermi energy not exactly at the nodal rings,
which leads to a finite density of states in the bulk and allows the screening of the localized charge.
Therefore, in order to measure and make use of the localized charge at domain walls directly in experiments,
we first need to prepare a TDSM with particle-hole symmetry in the bulk band structure,
which is an important problem to be solved in future material design and band calculations.

\subsection{Topology of nodal rings and drumhead states}
Finally, we verify the topological origin of the drumhead states obtained in our numerical calculations.
In order to discuss the emergence of in-gap states at the domain boundary,
here we focus on the topology of the 1D Hamiltonian $H_{k_y,k_z}(k_x)$ at the fixed transverse momentum $(k_y,k_z)$.
In case $J\neq 0$ and $J'=0$,
this system possesses the chiral symmetry $\Gamma = \tau_x \sigma_y$
for any direction of magnetization $\boldsymbol{n}$,
as seen in the previous section,
whereas there is no time-reversal or particle-hole symmetry.
Therefore, this 1D system belongs to the chiral unitary class AIII,
which is characterized by the topological invariant of $\mathbb{Z}$ \cite{Schnyder_2008,Schnyder_2009,Ryu_2010}.
This integer topological number corresponds to the winding number in the bulk,
and the number of zero modes at the boundary.
If we flip the magnetization from $\boldsymbol{n} = +\boldsymbol{e}_x$ to $-\boldsymbol{e}_x$ by time-reversal operation,
the winding number also changes its sign.
From this discussion,
we can regard the head-to-head domain wall as the boundary between two domains with different topological numbers $\pm 1$,
which should host twofold degenerate zero modes corresponding to the drumhead states at $E=0$ obtained above.

Indeed, the 1D Hamiltonian here can be mapped to the simple model of the 1D topological insulator of class AIII.
For clarity of discussion, we here focus on the behavior at the Dirac point, $k_y =0, k_z = k_{\mathrm{D}}$,
and take the magnetization $\boldsymbol{n} = (\pm 1,0,0)$,
under which the 1D Hamiltonian on lattice reads as a $4 \times 4$-matrix,
\begin{align}
    H_{0,k_{\mathrm{D}}}(k_x) = t \sin k_x a \ \tau_x \sigma_z + 2M_1 (1-\cos k_x a) \tau_z + J n_x \sigma_x.
\end{align}
Since this matrix commutes with $\tau_z \sigma_x$,
it can be decomposed into two subspaces labeled by the eigenvalues $\lambda = \pm$ of $\tau_z \sigma_x$,
\begin{align}
    H_{0,k_{\mathrm{D}}}^\lambda(k_x) &= \left[\lambda t \sin k_x a\right] s_y + \left[ \lambda J n_x + 2M_1 (1-\cos k_x a) \right] s_z,
\end{align}
where $s_{x,y,z}$ denotes the Pauli matrix in each subspace.
This $2 \times 2$-matrix form exactly corresponds to the minimal model (Jackiw--Rebbi model) of AIII topological insulator in 1D \cite{Jackiw_1976,Teo_2010},
whose winding number $N_\lambda$ in each sector $\lambda = \pm$ becomes
\begin{align}
    N_+ = \begin{cases}
        0 & (n_x =+1) \\
        -1 & (n_x =-1)
    \end{cases}
    , \quad
    N_- = \begin{cases}
        +1 & (n_x =+1) \\
        0 & (n_x =-1)
    \end{cases}
\end{align}
as long as $|J|<4M_1$.
As a consequence,
the net topological number $N= N_+ + N_-$ becomes
\begin{align}
    N = \begin{cases}
        +1 & (n_x =+1) \\
        -1 & (n_x =-1)
    \end{cases} .
\end{align}
By comparing the two domains of $n_x = +1$ and $-1$,
the net topological number $N$ is different by $2$,
and hence there arises two in-gap zero modes at their domain boundary.
Even if the transverse momentum $(k_y,k_z)$ is slightly off the Dirac point
or the magnetization $\boldsymbol{n}$ has $z$-component,
the topological number remains unchanged unless the 1D spectrum closes the bandgap.
Therefore, the net topological number is $\pm 1$ and flips its sign at the domain boundary
for any $(k_y,k_z)$-point inside the nodal rings,
which yields the doubly degenerate drumhead states inside the nodal rings.

We should note that the symmetry argument given here relies on the presence of chiral symmetry.
In the presence of $J' \neq 0$ in the exchange coupling term,
the chiral symmetry by $\Gamma = \tau_x \sigma_y$ is violated,
and hence there is no topological restriction on the electrons at the domain boundary.
As long as $|J'| \ll |J|$,
we can still expect that there remain a pair of localized state around the domain wall,
with their double degeneracy at $E=0$ split by $J'$ only perturbatively.

\section{Conclusion}
\label{sec:conclusion}

In the present article,
we have investigated the effect of local magnetic moments and magnetic textures on the electronic band structure of TDSM.
One of the central results in this work is 
the transmutation of the Dirac points into nodal rings or Weyl points
triggered by the magnetic order.
The nodal structure, either nodal rings or Weyl points,
depends on the direction of magnetization and the orbital \modify{dependence in} the exchange coupling between the local magnetic moments and the electron spins,
which is summarized in the form of phase diagrams as shown in Fig.~\ref{fig:phase-diagram}.
We can thus regard TDSM as a good candidate system to switch the Weyl semimetallic phase and the nodal-line phase by controlling the magnetization,
with external magnetic fields or current-induced torques.

Another important finding is the emergence of zero-energy boundary modes localized at magnetic domain walls.
Those localized modes form flat bands enclosed by the nodal rings,
which are similar to the drumhead surface states widely known in the context of nodal-line semimetals.
Unlike the mirror topological number of $\mathbb{Z}_2$ in nodal-line semimetals protected by mirror symmetry,
the drumhead boundary states in the magnetic nodal-line state found here are shown to correspond to the integer topological number of $\mathbb{Z}$ in the bulk.
We have also found that these drumhead states contribute to the electric charging of domain walls,
whose magnitude depends on the size of the nodal rings.
Provided that the localized charge is free from electrostatic screening by the bulk carriers,
which is satisfied if the bulk bands are particle-hole symmetric about the nodal rings,
we expect that the localized charge may be useful
in the electric manipulation of magnetic domain walls in spintronics devices,
as proposed in the context of magnetic topological insulators and Weyl semimetals as well \cite{Araki_2016,Araki_2018,Nomura_2011,Kurebayashi_2019,Kurebayashi_2017,Kim_2019,Araki_2021_THT}.

\begin{acknowledgment}
    The authors appreciate K.~Kobayashi and Y.~Ominato for fruitful discussions at Tohoku University.
    Y.~A. and K.~N. are supported by JSPS KAKENHI Grant Number 20H01830.
    Y.~A. is supported by the Leading Initiative for Excellent Young Researchers (LEADER).
    K.~N. is supported by JST CREST Grant No.~JPMJCR18T2.
\end{acknowledgment}

\end{document}